\input{aipcheck}

\documentclass[
    ,final            
  ]
  {aipproc}

\layoutstyle{6x9}

\begin{document}

\title{The circum-galactic gas around cosmologically simulated disks}


\author{St\'ephanie Courty}{
  address={Jeremiah Horrocks Institute for Astrophysics $\&$ Supercomputing, University of Central Lancashire, United Kingdom}
}

\author{Brad K. Gibson}{
  address={Jeremiah Horrocks Institute for Astrophysics $\&$ Supercomputing, University of Central Lancashire, United Kingdom}
}

\author{Romain Teyssier}{
  address={CEA Saclay, DSM/IRFU/SAp, France}
  ,altaddress={Institute for Theoretical Physics, University of Zurich, Switzerland} 
}

\begin{abstract}

We analyze the physical properties and infall rates of the
circum-galactic gas around disks obtained in multi-resolved,
cosmological, AMR simulations. At intermediate and low redshifts,
disks are embedded into an extended, hot, tenuous corona that
contributes largely in fueling the disk with non-enriched gas whereas
the accretion of enriched gas from tidal streams occurs throughout
episodic events. We derive an infall rate close to the disk of the
same value as the one of the star formation rate in the disk and its
temporal evolution as a function of galacto-centric radius nicely
shows that the growth of galactic disks proceeds according to an
inside-out formation scenario.

\end{abstract}

\maketitle


Several direct and indirect evidences strengthen the idea that
galactic disks are fueled at low redshift by external, accreted
gas. Regarding the Milky-Way for instance, these evidences come from
the disk itself as gas infall is required to sustain its moderate star
formation rate and is also needed to interpret the metal content
distribution in the solar neighborhood. But the physical conditions of
the accreted gas, as well as its form (diffuse and/or clumpy), are
still open issues. The many detections of highly ionized high velocity
clouds at high latitudes around the Galaxy with FUSE \cite{Sembach03}
have been interpreted as clumps of gas moving into a hot and tenuous
corona, as suggested by Spitzer \cite{Spitzer56} in the fifties,
though it would have a larger extent than the corona considered in
this seminal paper. Observational and theoretical works \cite{Peek08,
Bland-Hawthorn09} suggest that infall of cold gas clumps would be
insufficient to sustain the star formation rate in the Galactic disk
and that infall of additional hot material is needed.

Though much of the emphasis nowadays is focused on the so-called cold
accretion mode at high redshift, we here concentrate our discussion
over the last 10 Gyrs and attempt to connect the physical conditions
of the circum-galactic gas with the rate it falls onto the disk.
\\

We use multi-resolved, large-scale structures, N-body/hydrodynamical
simulations whose initial conditions are re-centered on a Milky-Way
sized halo (M$_{dyn}$=$7.2\times10^{11}$ M$_{\odot}$). Such a
multi-resolved approach allows to perform cosmological simulations of
given halos at a high spatial resolution. We use the AMR RAMSES
code \cite{Teyssier02} and a first set of our results is reported
in \cite{Sanchez-Blazquez09}. The simulations include, in addition to
gravitation and gas dynamics, star formation and its associated
thermal and kinetic feedback due to the explosions of the young
stellar populations, and follow the gas heavy element content. In the
central area, the mass resolution of the simulation is $6\times10^6$
M$_{\odot}$ for the dark matter particle mass and $10^6$ M$_{\odot}$
for the initial baryonic mass per cell. The coarse grid is 512$^3$ and
at the maximum level of refinement the spatial resolution is
20$h^{-1}$Mpc/$2^{16}$=436 pc.
\\

\begin{figure}
  \begin{minipage}[c]{0.5\textwidth} \includegraphics[height=.33\textheight,angle=270]{001scourty_fig1.ps} \end{minipage} \begin{minipage}[c]{0.5\textwidth} \includegraphics[height=.22\textheight]{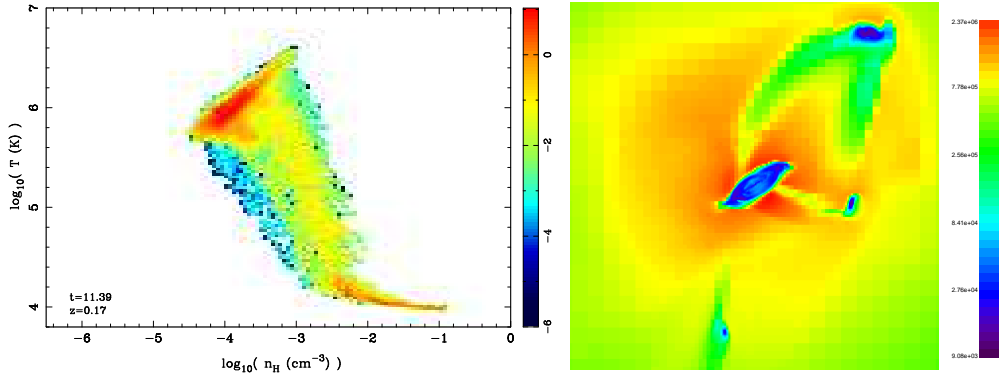} \end{minipage} 
\caption{Left panel: Density-temperature gas phase diagram of the gas within 60 kpc of the
  center of the galaxy at z=0.17 ; the gas in the disk and satellites
  is not accounting for. Right panel: Mass-weighted temperature map
  centered on the disk at the same redshift (the size of the image is
  70$\times$70 kpc and its depth is 70 kpc).}  \label{fig:map}
\end{figure}

Our simulated disk is embedded into a hot corona (right panel of
Fig.~\ref{fig:map}) with temperatures $T$ typically between $6.10^5$
and $1.5 \times 10^6$ K and densities $n_H$ higher than a few
$10^{-5}$ cm$^{-3}$, as shown by the phase diagram on the left
panel. The extent of the corona is much less than the virial radius of
the halo. The gas between the corona and the outer regions of the halo
presents lower densities ($n_H>10^{-6}$ cm$^{-3}$) and lower
temperatures ($T>10^5$ K), preventing this gas to cool on timescales
smaller than the Hubble time. The plume of gas present in the phase
diagram with T$<5.10^5$ K and $n_H$ higher than a few $10^{-4}$
cm$^{-3}$ betrays the change of conditions due to the interactions
between the corona and fast-moving satellites and to the tidal streams
resulting from close encounters between the disk and small companions
(the blue clumps in the temperature map). The bulk of the gas in the
corona has a metallicity of $10^{-2}$ in solar units whereas the gas
metallicity in the streams is typically a tenth of the solar
abundance. Within 60 kpc, the amount of hot gas is $2.5 \times 10^9$
M$_{\odot}$ ; \cite{Bregman-Lloyd-Davies07} estimate a similar amount
for the inner halo of the Milky-Way through X-Ray absorption
measurements.

\begin{figure}
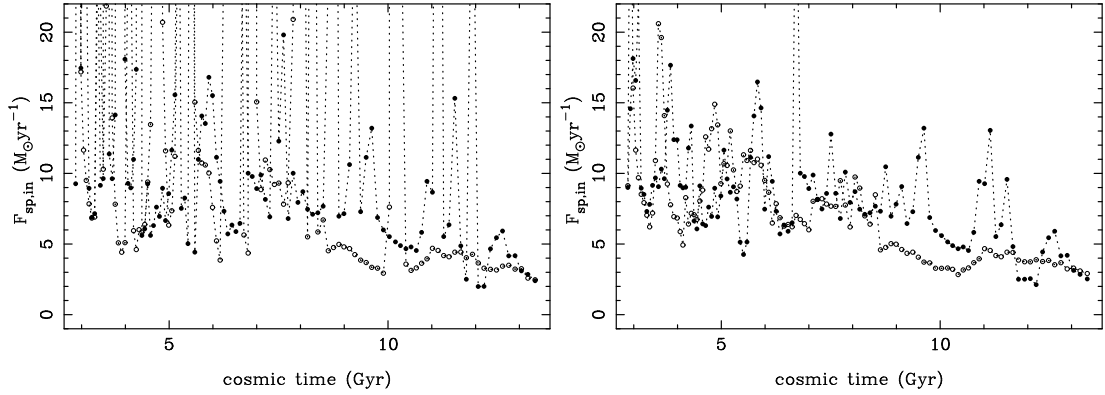

  \includegraphics[height=.33\textheight,angle=270]{001scourty_fig3.ps} \includegraphics[height=.33\textheight,angle=270]{001scourty_fig4.ps} \caption{Left
  panel: Flux of gas flowing in through spherical surfaces located at
  the virial radius (open dots) and at 30 kpc (filled dots) from the
  disk. Right panel: Same but without accounting for the gas within
  satellites.}  \label{fig:sphflux}
\end{figure}

\begin{figure}
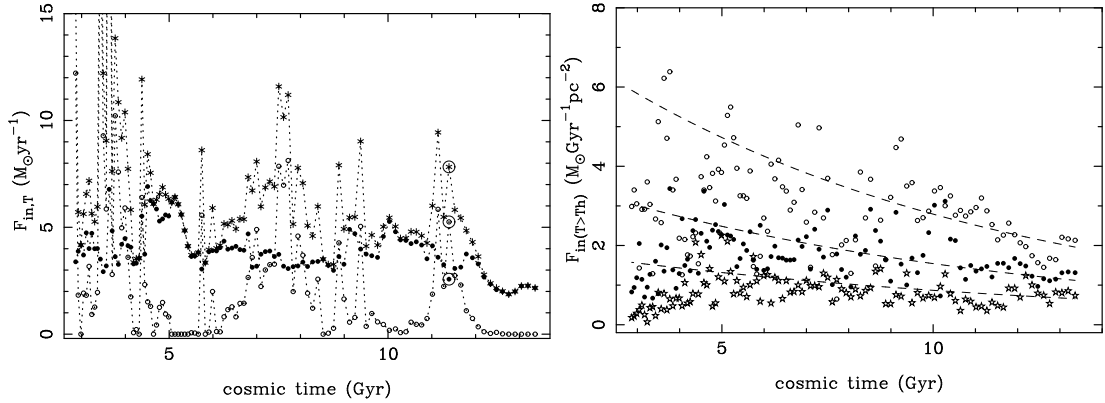

  \includegraphics[height=.33\textheight,angle=270]{001scourty_fig5.ps} \includegraphics[height=.33\textheight,angle=270]{001scourty_fig6.ps} \caption{Left
  panel: Flux of gas flowing in through parallel slabs of 17 kpc of
  radius located at 7 kpc above and below the equatorial plane of the
  disk (stars) ; the gas within satellites is not accounted for. The
  contributions from the hot ($T>10^6$ K) and warm/cold ($T<10^6$ K)
  gas is shown with the filled and open dots, respectively. The large
  circle on each curves indicate the redshift at which
  Fig.~\ref{fig:map} is displayed. Right panel: Same but considering
  the gas crossing the slabs at 3 given galacto-centric radii, 12.5
  (stars), 8.5 (filled dots), 4.5 (open dots) kpc, and accounting only
  for the hot gas (note that infall rates are now expressed per unit
  of area).}  \label{fig:flux}
\end{figure}

There is a large reservoir of gas in the halo to potentially fuel the
disk: At z=0 within the virial radius, while the disk accounts for
39$\%$ of the available gas and the gas within the ``virial'' radius
of satellites for 3$\%$, the rest (58$\%$) is part of the halo. But
only the gas in the corona, the close environment, presents the
appropriate physical conditions for it to be accreted onto the
disk. To quantify how much gas is accreted, we first compute the flux
of gas flowing in through spherical surfaces located either at the
virial radius or at 30 kpc of the center of the disk. The left panel
in Fig.~\ref{fig:sphflux} shows an irregular evolution whose peaks
correspond to satellites crossing the surfaces. Discounting the gas
within the satellites allows to retrieve a smoother evolution (right
panel), slowly decreasing with time over the last 10 Gyrs, but the
spherical flux at 30 kpc is still more disturbed than the one at the
virial radius. We note that the flux at z=0 is 3 M$_{\odot}$yr$^{-1}$,
the value of the star formation rate in our simulated disk.

To assess the infall rate as close as possible to the disk, we then
estimate the flux of gas flowing in through slabs parallel to the
equatorial plane and located at 7 kpc below and above it. This height
is high enough to avoid accounting for the recirculating gas, ejected
from the disk through outflows. The star symbols in the left panel in
Fig.~\ref{fig:flux} display the evolution of this infall rate where
the gas belonging to satellites has been discarded. The filled and
open symbols separate the contribution from the hot and cold gas,
respectively. We have also distinguished the gas according to its
metallicity. Overall we note that the smooth envelop of the infall
rate is composed of hot and metal-poor gas whereas the episodic events
are dominated by cold and metal-enriched gas, mainly coming from the
tidal streams ; a typical example of such an event is emphasized by
the circle marking the time-step at which the temperature map is shown
in Fig.~\ref{fig:map}.

If we now consider the gas flowing in through the same slabs as before
but estimate its infall rate at 3 given galacto-centric radii (12.5,
8.5, 4.5 kpc), we see a clear evolution of the infall rate as it
decreases with time and decreases with radius at a given time. This
result is even clearer if we only account for the hot gas (right panel
of Fig.~\ref{fig:flux}). Moreover the exponential time-scales
($\sim$9-10 Gyrs) are of the same orders as those involved in the
two-infall models \cite{Chiappini97}, a basic and important ingredient
of most of galactic chemical evolution models. This is an intriguing
result as it is here based on cosmological simulations in which
galactic disks form ab initio. Figure~\ref{fig:flux} is a nice
illustration of the inside-out formation scenario \cite{Larson76}
stating that the growth of galactic disks proceeds externally. It also
shows that the bulk of the gas involved in the two-infall model might
be dominated by the accretion of hot, metal-poor gas ; the function of
the tidal streams being of providing metal-enriched gas through a
warm/cold phase.
\\

Chemodynamical simulations with the RAMSES code are under way to allow
for comparisons with observed chemical abundance distributions in
disks and help understand the links between accreted gas and star
formation. Similar cosmological simulations at higher resolution will
provide a better description of the corona by possibly resolving its
interaction with fast moving gas clouds that could explain some of the
observed highly-ionized HVCs. What is happening in the circum-galactic
environment might be crucial in connecting the warm gas on the large
scales of the universe (inter-galactic medium/filaments) with the
accreted gas on the small, disk scales.
\\


\begin{theacknowledgments}

Simulations were carried out on COSMOS (the UK's National Cosmology
Supercomputer) and the University of Central Lancashire's High
Performance Computing Facility. The parent N-body simulation was
performed within the framework of the Horizon collaboration
(www.projet-horizon.fr).

\end{theacknowledgments}

\bibliographystyle{aipproc}   

\bibliography{001scourty_biblio}

\end{document}